\documentclass[aps,prl,twocolumn,
	groupedaddress,superscriptaddress,
	amsfonts,amssymb,amsmath,floatfix,
	citeautoscript]{revtex4-1}

\usepackage{graphicx}
\usepackage[centering,hmargin=20mm,tmargin=30mm,bmargin=25mm]{geometry}
\usepackage{multirow}
\usepackage{newtxtext}
\usepackage[cmintegrals]{newtxmath}

\usepackage{xcolor}
\usepackage{hyperref}
\hypersetup{colorlinks,
	linkcolor={blue!75!black!80!yellow},
	citecolor={blue!75!black!80!yellow},
	urlcolor={blue!75!black!80!yellow}
}

\makeatletter
\renewcommand\@make@capt@title[2]{%
	\@ifx@empty\float@link{\@firstofone}{\expandafter\href\expandafter{\float@link}}%
	\sffamily{\textbf{#1}}\@caption@fignum@sep#2
}%

\makeatother

\newcommand{\HarvardSEAS}{John A. Paulson School of Engineering and Applied Sciences, Harvard University, Cambridge, MA, USA}
\newcommand{\MITPhy}{Department of Physics, Massachusetts Institute of Technology, Cambridge, MA, USA}

\usepackage[normalem]{ulem}

\begin{document}

\title{Variational theory of non-relativistic quantum electrodynamics}

\author{Nicholas Rivera}\email{nrivera@seas.harvard.edu}\affiliation{\HarvardSEAS}\affiliation{\MITPhy}
\author{Johannes Flick}\email{flick@seas.harvard.edu}\affiliation{\HarvardSEAS}
\author{Prineha Narang}\email{prineha@seas.harvard.edu}\affiliation{\HarvardSEAS}

\date{\today}

\begin{abstract}
 The ability to achieve ultra-strong coupling between light and matter promises to bring about new means to control  material properties, new concepts for manipulating light at the atomic scale, and fundamentally new insights into quantum electrodynamics (QED). Thus, there is a need to develop quantitative theories of QED phenomena in complex electronic and photonic systems. In this Letter, we develop a variational theory of general non-relativistic QED systems of coupled light and matter. Essential to our ansatz is the notion of an effective photonic vacuum whose modes are different than the modes in the absence of light-matter coupling. This variational formulation leads to a set of general equations that can describe the ground state of multi-electron systems coupled to many photonic modes in real space. As a first step towards a new  \textit{ab initio} approach to ground and excited state energies in QED, we apply our ansatz to describe a multi-level emitter coupled to many optical modes, a system with no analytical solution. We find a compact semi-analytical formula which describes ground and excited state energies to less than 1\% error in all regimes of coupling parameters allowed by sum rules. Additionally, our formulation provides essentially a non-perturbative theory of Lamb shifts and Casimir-Polder forces, as well as suggesting new physical concepts such as the Casimir energy of a single atom in a cavity. Our method should give rise to highly accurate descriptions of phenomena in general QED systems, such as Casimir forces, Lamb shifts, spontaneous emission, and other fluctuational electrodynamical effects. 
\end{abstract}

\maketitle


Recent years have brought an explosion of progress in the study of light-matter interactions in the non-perturbative regime of quantum electrodynamics (QED)~\cite{flick7strong,ruggenthaler2017b,forn2018ultrastrong,baranov2018}. Ultra-strong, or even deep-strong coupling is now regularly observed in systems involving electromagnetic cavities coupled to superconducting qubits~\cite{blais2004,wallraff2004,yoshihara2017superconducting,forn2017ultrastrong}, large ensembles of molecules~\cite{hutchison2012,coles2014,coles2014b,shalabney2015coherent, thomas2016,ebbesen2016,stranius2018,thomas2018}, Landau level systems ~\cite{scalari2012ultrastrong,zhang2016collective},  quantum wells coupled to cavities ~\cite{todorov2010ultrastrong,geiser2012ultrastrong}, and even in few-molecule systems ~\cite{benz2016,chikkaraddy2016}. Proposals for new platforms of ultra-strong coupling include emitters coupling to highly confined polaritons in metals and polar insulators \cite{rivera2016shrinking}, heavy ions coupled to optical media via the Cerenkov effect \cite{carmes2018non}, and many more. The proposed applications for ultra- and deep-strong coupling of light and matter are similarly broad, including simulation of many-body systems~\cite{forn2018ultrastrong}, altering chemical reactivity~\cite{hutchison2012, thomas2016,thomas2018, flick2017, herrera2016,feist2017,martinez2017} and electronic transport properties~\cite{orgiu2015} and realizing analogues of nonlinear optical processes with vacuum fluctuations~\cite{kockum2017deterministic}. Concomitantly with these exciting experimental developments are also theoretical developments in the study of QED systems \textit{ab initio}. Through `reduced quantity theories' such as quantum electrodynamical density functional theory (QEDFT)~\cite{tokatly2013,ruggenthaler2014,flick2015,dimitrov2017,flick2018,flick2018b}, one is now able to calculate observables in large molecules coupled to realistic optical cavities~\cite{flick2017c, flick2018b,flick2018}. 

 In this Letter, we establish a variational framework to analyze complex light-matter systems from first principles. Although  \textit{ab initio} methods such as QEDFT are exact in principle and provide access to all observables, a number of practical difficulties arise related to: the lack of simple exchange-correlation functionals to describe the ground state energy, as well as other more involved observables, the difficulty of obtaining real-space information about the photons as they are affected by light-matter coupling, the difficulty of handling excited state energies, and the common use of the long-wavelength (dipole) approximation. A variational framework, as we shall show, flexibly allows a real-space description of the electrons and photons as they are modified by the coupling and also beyond the dipole approximation. Beyond these advantages, a variational framework also allows conceptual insights, as we shall show, into a simple non-perturbative theory of Lamb shifts, into a quasiparticle description of QED systems, and into the notion of Casimir forces in the limit of one atom. A variational framework also allows compact semi-analytical formulae to describe complex systems which may assist the development of functionals for use in QEDFT. 

Motivated by all of these potential advantages, we now develop an ansatz in which the ground state can be considered as a factorizable state of effective matter and effective photon quasiparticles, both in their respective vacuum states. This ansatz $-$ reminiscent to, but qualitatively distinct from, the Hartree-Fock ansatz~\cite{szabo1989} of electronic structure theory $-$ leads to coupled eigen-equations describing ground and excited states of the light-matter system. We apply our ansatz to describe ground and excited states in a multi-level emitter coupled to many photonic modes. We find that for light-matter couplings that respect sum rules, our method yields ground and excited state energies to a remarkable accuracy of up to 99\%, even in deeply non-perturbative coupling regimes.  In regimes where our results are accurate, we have found the effective quasiparticle description of the ground state of QED.  Our findings also furnish a non-perturbative theory of the position-dependent energy (Lamb) shifts of ground and excited states that give rise to Casimir-Polder forces.



The QED Hamiltonian is given by $H = H_{mat}+H_{em}+H_{int}$ where $H_{mat}$ describes the matter in the absence of the quantized electromagnetic field, $H_{em}$ describes the photons in the absence of the matter, and $H_{int}$ describes the coupling between light and matter. The matter Hamiltonian takes the form:
\begin{align}
H_{el} &= \int d^3x ~\psi^{\dagger}(\mathbf{x})\left(-\frac{\hbar^2\nabla^2}{2m} + v_{ext}(\mathbf{x}) \right)\psi(\mathbf{x}) \nonumber \\ &+ \frac{1}{2}\int d^3x d^3x'~ \psi^{\dagger}(\mathbf{x})\psi^{\dagger}(\mathbf{x}')V(\mathbf{x}-\mathbf{x}')\psi(\mathbf{x}')\psi(\mathbf{x}),
\end{align}
where $v_{ext}$ is the one-body external potential, $V(\mathbf{x}-\mathbf{x}')$ is the two-body interaction kernel, and $\psi$ is the second-quantized electron field. 
\begin{figure}[t]
\includegraphics[width=8.5cm]{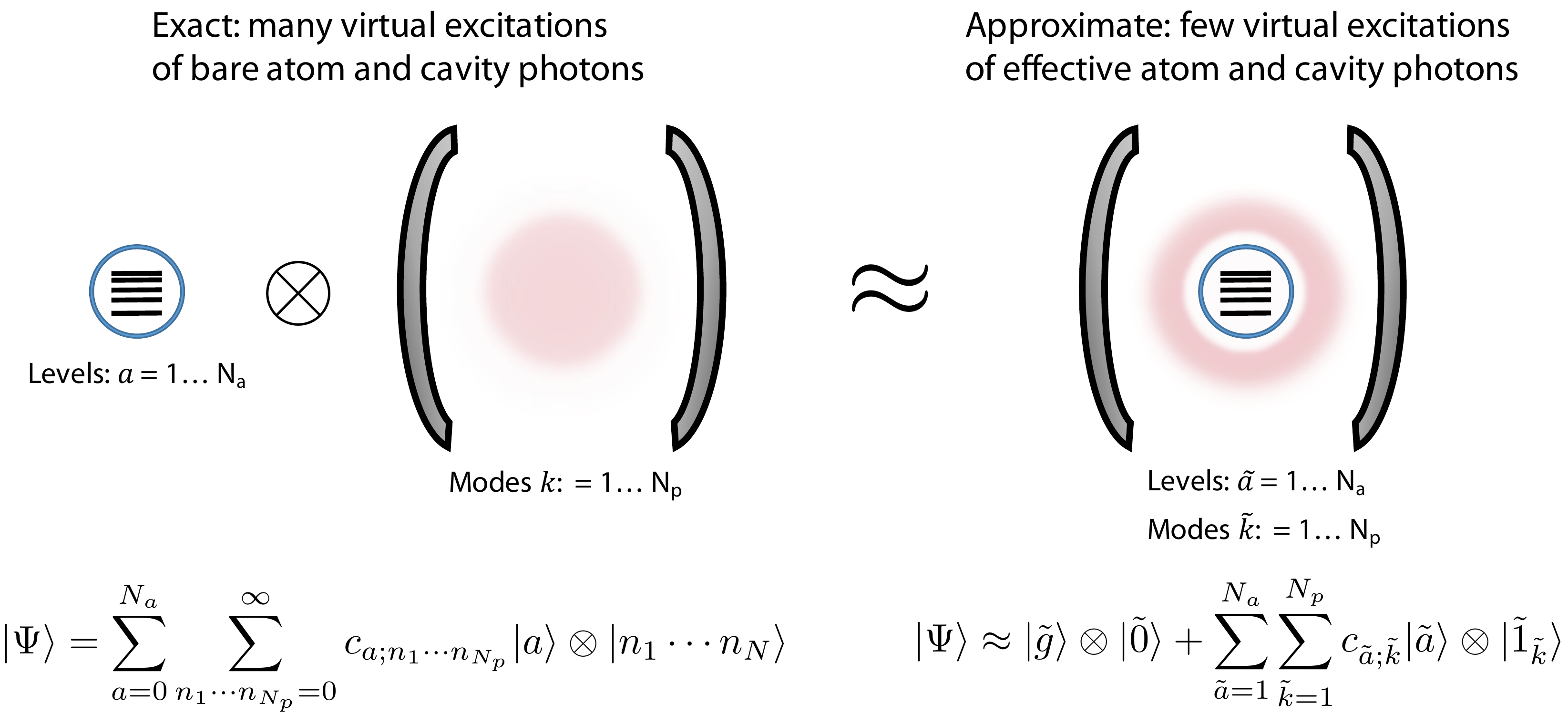}
\caption{\textbf{Ground-state ansatz applied to matter in a cavity: effectively decoupled matter and photons.} (Left) Bare description of the coupled light-matter ground state in terms of many virtual excitations of the emitter state and the bare cavity photons. (Right) Quasiparticle description of the coupled system as a factorizable state of an effective emitter in its ground state and the vacuum of an effective photonic degree of freedom.}
\label{fig:ansatz}
\end{figure}
Parameterizing the electromagnetic field purely in terms of a vector potential: $\mathbf{E} = -\partial_t\mathbf{A}$ and $\mathbf{B} = \nabla\times\mathbf{A}$ renders the free electromagnetic Hamiltonian as
\begin{equation}
H_{em} = \frac{\epsilon_0}{2}\int d^3x~ \epsilon (\partial_t \mathbf{A}(\mathbf{x}))^2 + \mathbf{A}(\mathbf{x})\cdot(\nabla\times\mu^{-1}\nabla\times\mathbf{A}(\mathbf{x})),
\end{equation}
where $\epsilon$ and $\mu$ represent a non-dispersive and positive dielectric and magnetic background that the matter and photon occupy. For cases we consider in this work, these will be taken to be unity.

The interaction Hamiltonian takes the form:
\begin{align}
H_{int} &= \frac{-i\hbar e}{2m}\int d^3x ~\psi^{\dagger}(\mathbf{x})(\mathbf{A}(\mathbf{x})\cdot\nabla +  \nabla \cdot \mathbf{A}(\mathbf{x}))\psi(\mathbf{x}) \nonumber \\ &+ \frac{e^2}{2m}\int d^3x ~\psi^{\dagger}(\mathbf{x})\psi(\mathbf{x})\mathbf{A}^2(\mathbf{x}).
\end{align}

The full Hamiltonian $H$, which depends on the fields $\psi$ and $\mathbf{A}$ can be parameterized in terms of an orthonormal set of electron single-particle wavefunctions (orbitals) $\{\psi_n\}$, and in terms of a set of photonic mode functions (orbitals) $\{\mathbf{F}_i\}$. The electron field operator takes the form $\psi(\mathbf{x}) = \sum_n \psi_n(\mathbf{x})c_n$.
The $c_n$ is an annihilation operator for an electron corresponding to state $n$. The electromagnetic field operator takes the form $\mathbf{A}(\mathbf{x}) = \sum_i\sqrt{\frac{\hbar}{2\epsilon_0\omega_i}} \left(\mathbf{F}_i(\mathbf{x})a_i+\mathbf{F}^*_i(\mathbf{x})a^{\dagger}_i\right)$, where the $a_i^{(\dagger)}$ annihilate (create) a photon in mode $i$. In the electromagnetic field operator, we parameterize not only by the mode functions but also by mode frequencies. The normalization chosen for the electron wavefunctions is $\int d^3x~ \psi_m^*\psi_n = \delta_{mn}$ while for the photon mode functions, it is $\int d^3x~\epsilon\mathbf{F}_i^*\cdot\mathbf{F}_j = \delta_{ij}$ \cite{joannopoulos2011photonic}.

With the Hamiltonian described, we now move to develop a variational theory of the ground state. In the variational theorem, we choose an ansatz $|\Omega\rangle$ for the ground state of $H$. The variational theorem ensures that $\langle \Omega|H|\Omega\rangle$ is an upper bound for the ground state energy. Parameterizing the ground state to generate a family of ground states, and minimizing $\langle \Omega|H|\Omega\rangle$ with respect to the introduced parameters gives the best upper bound for the ground state energy for the chosen family of ground states.  We choose as our ansatz
\begin{equation}
|\Omega\rangle = \left( \prod\limits_n c_n^{\dagger}|0_n\rangle\right) \otimes \left(\bigotimes_i|0_i\rangle\right).
\label{eq:ansatz}
\end{equation}
In such an ansatz, $\prod\limits_n c_n^{\dagger}|0_n\rangle$ represents a "filled Fermi sea" for effectively non-interacting electrons, and $\left(\bigotimes_i|0_i\rangle\right)$ represents a "photonic vacuum" for effectively non-interacting photons (see Fig. ~\ref{fig:ansatz}). Implicitly, this ansatz, once we take the expectation value $\langle \Omega|H|\Omega\rangle$, denotes a family of ansatzes labeled by all the possibilities for the electron wavefunctions,  photon mode functions, and photon mode frequencies. Thus, we shall minimize the expectation value with respect to $\psi_n, \psi_n^*, \mathbf{F}_i, \mathbf{F}_i^*$, and $\omega_i$.  We enforce that the matter and photon remain normalized by constructing the Lagrange function:
\begin{align}
&\mathcal{L}[\{ \psi_n,\psi_n^* \},\{ \mathbf{F}_i,\mathbf{F}_i^*,\omega_i \}] = \langle \Omega |H|\Omega\rangle \\
&- \sum_n E_n\left(\int d^3x ~\psi_n^*\psi_n - 1 \right) - \sum_n \frac{\hbar\lambda_i}{2}\left(\int d^3x ~\epsilon\mathbf{F}_i^*\cdot\mathbf{F}_i - 1 \right),\nonumber
\end{align}
with the $E_n$ and $\frac{\hbar\lambda_i}{2}$ being the Lagrange multipliers that enforce the normalization conditions. Evaluating the expectation value of the Hamiltonian, and minimizing the Lagrange function  immediately yields:
\begin{align}
&\left(\frac{\mathbf{p}^2}{2m}+v_{ext}(\mathbf{x}) \right)\psi_i(\mathbf{x}) + \nonumber \\ &\sum\limits_{j=1}^N \int d^3x' ~ V(\mathbf{x}-\mathbf{x}')\psi^*_j(\mathbf{x}')\psi_j(\mathbf{x}')\psi_i(\mathbf{x}) \nonumber \\ & - \sum\limits_{j=1}^N \int d^3x' ~ V(\mathbf{x}-\mathbf{x}')\psi^*_j(\mathbf{x}')\psi_j(\mathbf{x})\psi_i(\mathbf{x}')  \nonumber \\ &+ \frac{\hbar e^2}{4m\epsilon_0}\left(\sum_n \frac{1}{\omega_n}|\mathbf{F}_n|^2\right)\psi_i(\mathbf{x})   = E_i\psi_i(\mathbf{x}),
\label{eq:mhf-electron}
\end{align}
for the electron orbitals and energies. We see that in addition to the one-body and Hartree-Fock terms for the electrons, the effect of the QED coupling is to add a one-body ponderomotive potential. 

For the photon orbitals and energies, the minimization yields:
\begin{equation}
\left( \nabla\times\nabla\times - \frac{\omega_i^2}{c^2}\left(1-\frac{\omega_p^2(\mathbf{x})}{\omega_i^2} \right)\right)\mathbf{F}_i = 0,
\label{eq:mhf-photon}
\end{equation}
where $\omega_p^2(\mathbf{x}) = \frac{e^2}{m\epsilon_0}\sum\limits_{n=1}^N |\psi_n(\mathbf{x})|^2$ is a position-dependent squared-plasma frequency which will push the photon orbitals out of the region where the emitter is located. Equations (\ref{eq:mhf-electron}) and (\ref{eq:mhf-photon}) are main results and can be used to describe ultra-strongly coupled systems in three dimensions, in an arbitrary photonic system, and with multi-electron matter. Excited states in this framework can be identified with matter and photon quasiparticle excitations.

Immediately, we notice that the term in the interaction Hamiltonian linear in the vector potential (the "$A\cdot p$ term") makes no contribution to the expectation value of the ground state of the energy in this ansatz. At second order in the $A\cdot p$ term, virtual photon processes arise, such as Lamb shifts, whose emitter-position-dependence gives rise to van der Waals and Casimir-Polder forces  \cite{scheel2009macroscopic}. Thus, we seek to capture the effect of this term. Physically, this term will mix the factorizable ground state of Eq.~(\ref{eq:ansatz}) with states that simultaneously have virtual excitations of the matter and the electromagnetic field. The resulting state is now non-factorizable and we thus conclude that the term in the Hamiltonian linear in the vector potential leads to \textit{correlations} in the system, and contributes wholly at lowest order to the correlation energy of QED ground and excited states. We note that correlations can also be included in the energy shifts of excited emitter states, as well as states that already have photonic excitations \footnote{We briefly note that this behavior of the $A\cdot p$ and $A^2$-term is similar to the ${r}\cdot D$ and $r^2$ term in the length-gauge reported in recent work on the optimized effective potential~\cite{pellegrini2015,flick2017c} method for QEDFT including one-photon processes. }.
\begin{figure}[t]
\includegraphics[width=8.5cm]{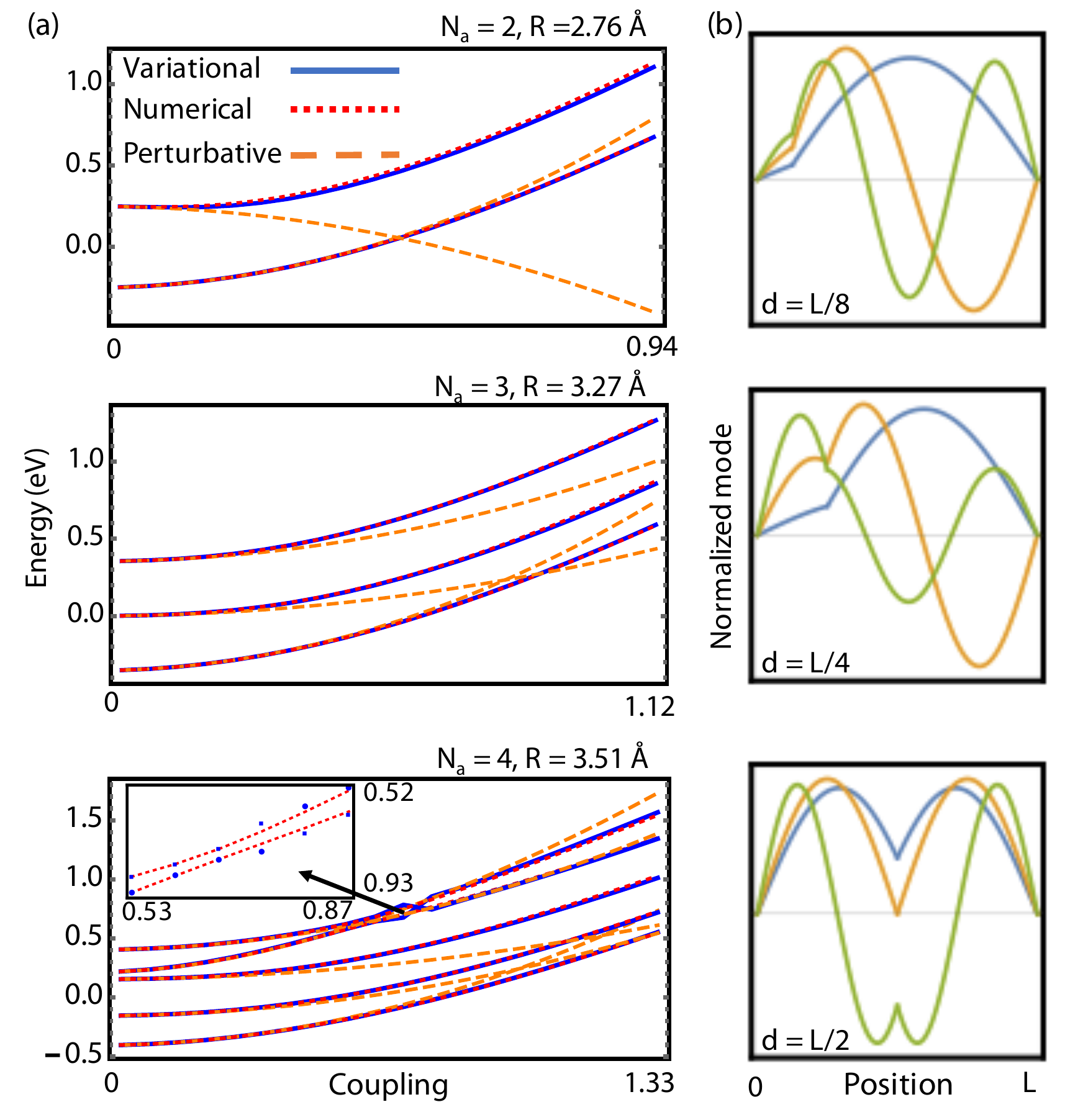}
\caption{\textbf{Variational theory of ground and excited states in the ultra-strong coupling regime of QED.} (a) Lowest few energy levels of a two (top), three (middle), and four (bottom) level system embedded in the middle of a one-dimensional cavity. The results of our variational method (blue) are compared to perturbation theory (orange), as well as numerical diagonalization (red) with the Fock space truncated to fifty cavity modes with no more than four photons. (Inset) Fourth and fifth energy levels shows a weak anti-crossing behavior which is well-reproduced by the variational theory. Blue denotes variational while red denotes numerical. (b) Mechanism of overestimation of couplings and resonances in perturbation theory: the modes derived from the variational theorem are always suppressed in the vicinity of the emitter, and this self-consistently decreases the coupling between the emitter and the field.  }
\label{fig:results}
\end{figure}

We capture the effect of correlations perturbatively. For the example of the ground state, we consider the second-order correction $\delta E$ to the ground state energy arising from the term in the Hamiltonian linear in the vector potential. That correction is given by 
\begin{equation}
\delta E = \frac{e^2\hbar^2}{8m^2\epsilon_0}\sum\limits_{i=1}^{\infty}\sum_{n=N_{\sigma}+1}^{\infty}\sum\limits_{m=1}^{N_{\sigma}} \frac{\Big| \int d^3x~\mathbf{F}_i^*\cdot\mathbf{j}_{nm}\Big|^2}{\omega_i(\omega_{mn} -\omega_i)},
\label{eq:e-shift}
\end{equation}
where $\mathbf{j}_{nm} = \psi^*_n\nabla\psi_m - (\nabla\psi^*_n)\psi_m$, $\omega_{mn} = \omega_m - \omega_n$, $N_{\sigma}$ is the number of occupied orbitals, equal to the number of electrons (divided by 2 if spin is retained).  In a method without self-consistency, the electron and photon orbitals and eigenvalues are those obtained from Eqs. (\ref{eq:mhf-electron}) and (\ref{eq:mhf-photon}), and then the electron energies and orbitals as well as the photon frequencies and orbitals, are plugged into Eq. (\ref{eq:e-shift}). By taking $m$ as an ansatz for an excited state, correlation corrections to excited states can also be found.

In what follows, we provide a proof-of-concept demonstration of the accuracy and content of the variational theory derived here. We consider the QED Hamiltonian corresponding to a single emitter placed at position $z=d$ in a one-dimensional cavity whose axis is along the $z$-direction. As the cavity is considered for simplicity to be one-dimensional, the electric field is oriented along a single direction, denoted $x$, while the magnetic field is oriented along a direction transverse to both the electric field and the cavity length, denoted $y$. Working under the long-wavelength (dipole) approximation, the Hamiltonian can then be written as:
\begin{equation}\label{eq:hamiltonian}
H = H_{\text{matter}}+\frac{\epsilon_0S}{2}\int dz~(E^2+c^2B^2)+\frac{q}{m}A(d)p + \frac{q^2}{2m}A^2(d),
\end{equation}
with the emitter charge now expressed as $q$, $E, B, A$ being the electric field, magnetic field, and vector potential, and $S$ being a normalization area of the cavity in the $xy$ plane. The fields can be expressed as a mode expansion, where for a cavity of length $L$, the modes are given by $F_n(z) = \sqrt{\frac{2}{L}}\sin\left(\frac{n\pi z}{L} \right)$ and the corresponding mode frequencies are $\omega_n = \frac{n\pi c}{L}$. The matter Hamiltonian we take to be a multilevel system with $N_a$ levels. The matter system we describe can thus be mapped to an $N_a$ site system, which be considered as a simplified model of a molecule within a tight-binding description. Thus we parameterize the general family of matter Hamiltonians as:
\begin{equation}\label{eq:matter_hamiltonian}
H_{\text{matter}} = \sum\limits_{i=1}^{{N_a-1}} V_i|i\rangle\langle i|+t(|i\rangle\langle i+1|+|i+1\rangle\langle i|).
\end{equation}
The momentum operator, we write as 
\begin{equation}\label{eq:momentum_operator}
p = \frac{-i\hbar}{R}\sum\limits_{i=1}^{N_a-1} \left(|i\rangle\langle i+1|-|i+1\rangle\langle i| \right),
\end{equation}
where $R$ is a constant with units of length representing roughly the difference in positions between sites. This physical interpretation however is rough: it is also a function of the hopping elements $t$, because we choose $R$ in this work such that the Thomas-Reiche-Kuhn (TRK) sum rule is enforced. In other words: $\frac{2}{m}\sum\limits_{i=2}^{N_a}\frac{|p_{ig}|^2}{E_i - E_a} = 1$, where $p_{ig} = \langle i|p|g\rangle$ are momentum matrix elements between different matter states.  Since the TRK is based on a full electronic real-space description, this sum rule does not rigorously apply to a discrete-level system. However, the matrix elements and energy levels of a few-level approximated Hamiltonian  are derived from an underlying real-space (infinite dimensional) Hamiltonian. Thus, a discrete system which has  $\frac{2}{m}\sum\limits_{i=2}^{N_a}\frac{|p_{ig}|^2}{E_i - E_a} > 1$ cannot exist physically. It thus places a bound on how strong the effect of the $A\cdot p$ term can be. The net effect is that the value of $R$ we choose is on the order of $\sqrt{\frac{\hbar}{2mt}}$. These considerations also imply that when we plot observables as a function of parameter, for fixed $R$, we vary the coupling by varying some external sum-rule independent measure such as the charge of the emitter, or the number of emitters collectively coupled to the mode. We choose the former.

The detailed derivations of the energies of states via the formalism introduced here are shown in the Supplementary Materials (SM). Here, we state the main results. Using an essentially one-dimensional version of Eq. (\ref{eq:mhf-electron}) and (\ref{eq:mhf-photon}), we calculate the electron orbitals, photon orbitals, and photon frequencies in the absence of correlations. In the absence of correlations, we found for example that the energy of any matter state $a$ with no photonic quasiparticles is given by:
\begin{equation}
E_{a} = E^{0}_a + \frac{1}{2}\sum\limits_{n=1}^{\infty}(\hbar\omega_n - \hbar\omega_n^0),
\label{eq:casimir}
\end{equation}
where $E_a^{(0)}$ is the energy of the matter state in the absence of coupling, $\omega_n$ are found in our framework, $\omega_n^0 = \frac{n\pi c}{L}$. Imposing continuity of the modes and discontinuity of their derivatives at $z=d$, the modes found in our framework have their frequencies given by
\begin{equation}
\cot\left(\frac{\omega_n}{c}d \right)+\cot\left(\frac{\omega_n}{c}(L-d) \right) = -\frac{q^2}{m\epsilon_0\omega_nc}.
\label{eq:cot}
\end{equation}
The corresponding "interacting" field mode profiles are given by
\begin{align}\label{eq:field_mode}
N_n^{-1}F_n(z) =~ &\theta(z-d) \left(\frac{\sin\left(\frac{\omega_nL}{c}\right)\sin\left(\frac{\omega_nd}{c}\right)\cos\left(\frac{\omega_nz}{c}\right)}{\sin\left(\frac{\omega_n(L-d)}{c}\right)}\right) \nonumber \\ 
-&\theta (z-d) \left(\frac{\cos\left(\frac{\omega_nL}{c}\right)\sin\left(\frac{\omega_nd}{c}\right)\sin\left(\frac{\omega_nz}{c}\right)}{\sin\left(\frac{\omega_n(L-d)}{c}\right)}\right) \nonumber \\ 
+&\theta (d-z) \sin\left(\frac{\omega_n z}{c} \right),
\end{align}
with the normalization constant
\begin{equation}\label{eq:mode_normalization}
N_n = 2\sqrt{\frac{1}{\frac{c}{\omega_n}\left(\frac{\omega_nL}{c}-\sin\left(\frac{\omega_nL}{c}\right) \right)\left(1+\frac{\sin^2\left(\frac{\omega_nd}{c}\right)}{\sin^2\left(\frac{\omega_n(L-d)}{c}\right)} \right)}}.
\end{equation}
The result of Eq. (\ref{eq:casimir})  says that in the absence of correlations, the energy of the system is the Casimir energy of the system. In particular, it has long been known that when two conducting plates are placed near each other, there is a Casimir energy associated with the fact that the zero-point energy of the nearby plates is different than the zero-point energy of plates infinitely apart. This is because the electromagnetic mode structure of two nearby plates is different from that of two infinitely separated plates. This Casimir energy is simply the difference between the interacting and non-interacting zero-point energies \cite{casimir1948attraction,lifshitz1956theory}. This logic can be applied to any arrangement of macroscopic polarizable objects. What is notable about the result of Eq. (\ref{eq:casimir}) is that our result says that the same logic about zero-point energy-differences can be applied to find the interaction energy case of a \textit{single atom} placed near a cavity, even though a single emitter is very far from the limit of a macroscopic polarizable object. 

In the presence of correlations  we must add to the energy a contribution of the form of Eq. (\ref{eq:e-shift}), specialized to the case of an emitter in a one-dimensional cavity. The interaction energy, given by Eqs. (\ref{eq:e-shift}) and (\ref{eq:casimir}) is semi-analytical once the bare emitter states are known, as it is fully specified by Eqs. (\ref{eq:cot}-\ref{eq:mode_normalization}) once the transcendental equation of Eq. (\ref{eq:cot}) is solved. We also apply the correlation correction to excited states as well, by using the second-order perturbation theory formula for the energy shift of excited states due to the $A \cdot p$ term, using the same electron and photon orbitals and frequencies as derived in Eqs. (\ref{eq:mhf-electron}) and (\ref{eq:mhf-photon}). In Fig. ~\ref{fig:results}(a), we show the result of this procedure when applied to calculate ground- and excited- state energies for two-, three-, and four-level systems coupled to a one-dimensional cavity. The relevant parameters for Fig. 2(a) are listed in the SM. In all cases, the agreement between our variational approach and numerical diagonalization is excellent, suggesting that our variational method is sufficiently flexible to capture ground states and excited states both with and without photonic excitations. This is to be contrasted with perturbation theory in the bare matter and photon states, which can both strongly over- and underestimate the energies. The most interesting case of disagreement arises in the case of the two-level system (top panel). For the two-level system considered here, the variational result agrees very well with numerical diagonalization, while perturbation theory predicts an energy which evolves with coupling in the wrong direction and is off from the true energy by over 100\%.

Importantly, the reason perturbation theory fails for first excited state, much more so than for the ground state, is that the first bare cavity mode is nearly resonant with the transition between ground and excited emitter states, leading to a very large negative contribution from the $A\cdot p$ of nearly $2$ eV, which is far larger than the spacing of the bare emitter levels. On the other hand, the variational estimate from our formalism finds no such large negative energy shift, and leads to an energy gap between the first two levels which is similar to the bare gap, and in agreement with numerical diagonalization. The reason for this behavior is that the effect of the plasma term in Eq. (\ref{eq:mhf-photon}) is to blue-shift all of the photon frequencies. In particular, for the largest coupling considered in Figure~\ref{fig:results}, we find that the lowest photon frequency is shifted to 0.99 eV, and then becomes far off-resonance from the bare emitter transition. The plasma term, as shown in Fig.~\ref{fig:results}b, also strongly reduces the coupling between light and matter by a different mechanism in which the field modes obtained from Equation (\ref{eq:mhf-photon}) are screened out of the emitter, thus self-consistently reducing the strength of the coupling between matter and field and the magnitude of the correlation term, as per Equation (\ref{eq:e-shift}). This is a so-called light-matter decoupling effect \cite{liberato2014}. The results of Fig. ~\ref{fig:results} very clearly demonstrates not only the accuracy of our ansatz, but provides insight into the mechanisms by which light-matter coupling saturates in the nonperturbative QED regime. 

Our results also demonstrate a non-perturbative theory of the Lamb shift and consequently Casimir-Polder forces. In particular, it is long known that energy levels of emitters can shift as a result of virtual photon emission and re-absorption. These energy shifts, called Lamb shifts, depend on the particular position of the emitter in the photonic structure it is embedded in. These shifts not only lead to changes in the transition frequencies of the emitter, but the position dependence of these energy shifts also implies forces on the emitter, typically called Casimir-Polder forces. Such Casimir-Polder forces are often calculated using the celebrated Lifshitz theory \cite{lifshitz1956theory}, which is equivalent to a derivation that applies of second-order perturbation theory in the form of Eq. (\ref{eq:e-shift}) using bare atomic and photonic properties \cite{scheel2009macroscopic}. Thus our calculation of the energy shifts via Eq. (\ref{eq:e-shift}), which uses the interacting photon modes and frequencies (Eqs. (\ref{eq:cot}-\ref{eq:mode_normalization})), which differ greatly from the bare modes and frequencies in the non-perturbative regime, provide a compact, relatively simple, and semi-analytical extension of the theory of Lamb shifts and Casimir-Polder forces to the non-perturbative regime.

With the advent of ultra-strong coupling and deep-strong coupling in QED systems, the theory posed here, when applied to more complex systems, could form the basis for understanding Casimir phenomena in the ultra-strong coupling regime. Additionally, the results developed here could be extended to matter  or photon systems with infinitely many degrees of freedom, also allowing to capture effects like spontaneous emission in the non-perturbative regime. Finally, one could use the non-perturbative real-space knowledge provided by the variational theory of how matter affects photons in order to design a photonic mode atom-by-atom.

\section{Acknowledgements}
We thank Prof. Joel Yuen-Zhou (University of California San Diego), Prof. Ido Kaminer (Technion Israel Institute of Technology), Prof. Marin Solja\v{c}i\'{c} (Massachusetts Institute of Technology) and Prof. John D. Joannopoulos (Massachusetts Institute of Technology), for useful discussions. N. R. recognizes the support of the DOE Computational Science Graduate Fellowship (CSGF) fellowship no.  DE-FG02-97ER25308. J. F. acknowledges financial support from the Deutsche Forschungsgemeinschaft (DFG Forschungsstipendium FL 997/1-1). This work was supported by the DOE Photonics at Thermodynamic Limits Energy Frontier Research Center under grant no. DE-SC0019140.


\bibliographystyle{apsrev4-1}
\bibliography{references}

\end{document}